\documentclass[intlimits,twoside,a4paper]{article}

\usepackage{amsmath,amssymb} %%%% already in cmpj3.sty
\usepackage{graphicx}        %%%% already in cmpj3.sty

\usepackage{bm}              %%%% place it below \usepackage{cmpj3}

\usepackage[cp1251]{inputenc}

\usepackage[eqsecnum]{cmpj3}

\issue{2019}{22}{2}{23702}
\doinumber{10.5488/CMP.22.23702}

\title{First principle investigation of hydrogen behavior in M doped Cu$_2$O (M $=$ Na, Li and Ti)%
}
\author[A. Larabi, A. Mahmoudi, M. Mebarki, M. Dergal]{A. Larabi\refaddr{1}\thanks{Corresponding author: A. Larabi, amina.larabi8@gmail.com, amina.Larabi@crtse.dz.
}, A. Mahmoudi\refaddr{2}, M. Mebarki\refaddr{1}, M. Dergal\refaddr{2}}
\addresses{
\addr{1} Centre de Recherche en Technologie des Semi-conducteurs pour l’Energ\'etique (CRTSE), 2, Bd Frantz Fanon, BP 140 Alger 7-Merveilles 16038, Algerie
\addr{2} Division Etude et Pr\'ediction des Mat\'eriaux, Unit\'e de Recherche Mat\'eriaux et Energies Renouvelables, LEPM-URMER, Universit\'e de Tlemcen, Algerie
}

\date{Received October 30, 2018, in final form March 4, 2019}

\begin{document}

\maketitle

\begin{abstract}
We study the hydrogen effect on the electronic, magnetic and optical properties of Cu$_2$O in presence of different dopants (Na, Li and Ti). The electronic properties calculations show that hydrogen changes the conductivity of Cu$_2$O from p to n-type. The results show that interstitial hydrogen atom prefers to locate in the tetrahedral site in Cu$_2$O system and it decreases the band gap value of the later. The Na or Li doping Cu$_2$O preserves the p-type conductivity of Cu$_2$O, while hydrogen is the source of n-type conductivity in Na or Li doped Cu$_2$O systems.  Ti doping increases the band gap value of Cu$_2$O and makes it an n-type semiconductor. Hydrogen increases the optical transmittance of M doped Cu$_2$O.
\keywords Cu$_2$O, hydrogen, density functional theory
\pacs  70.
\end{abstract}

\section{Introduction}
Owing to high optical absorption coefficient, non-toxicity, abundancy and inexpensive layer formation, copper oxide (Cu$_x$O) is appealing to a renewed and rising interest. It could have a potential role in fabricating semiconductor devices, among others, for nanoelectronic, spintronic and photovoltaic appli\-cations \cite{Ols79,Rak86,Tot60}. We know copper oxide as a p-type semiconductor in two stable forms, cupric oxide (CuO) and cuprous oxide (Cu$_2$O) and this latter enables higher conversion efficiency when used as an absorber material for solar cell applications \cite{O'ke61,McK67}. Moreover, the possibility of synthesizing nanoparticles enables excellent properties and opens the way to a new class of innovative and more efficient devices~\cite{Sag13}. The ability to control the conductivity type can be important for the synthesizing of Cu$_2$O thin films paired with other p-type materials. The challenge of p-type doping of Cu$_2$O must take hydrogen into account, since it is known to passivate any acceptors in solar cell absorbers \cite{Kil03} and reverse the polarity of conductivity in other materials \cite{Ott01}. Hydrogen is present in all semiconductors, including oxides, and is highly reactive, forming a complex with most impurities and defects in the lattice. Hydrogen is a potential source of n-type conductivity \cite{Kol09,Fig11,Jan07}. The capability of hydrogen to remove or stimulate the electrical and optical activity of impurities makes it sometimes a helpful, sometimes a disagreeable agent, in either case. However, a theoretical understanding of the principal interactions is very important. The diversity and complexity of hydrogen states, inside the lattice and in traps, has made their characterization impossible by experimental methods alone; the complementary role of theory is proved to be essential \cite{Bih09}. The earlier calculations suggested that interstitial hydrogen prefers to locate in the bond center site (BC) in CdTe and ZnO lattices \cite{Van00,Lar18,Rak09,Lar16}. It could locate in the tetrahedral site in other oxides such SnO$_2$ and TiO$_2$ \cite{Var09,Fil09}. Theoretical prediction and spectroscopy confirmed that hydrogen in semiconductors may be a cause of n-type conductivity; it forms shallow-donor states in the zinc oxide (ZnO) \cite{Van00,Cox01,Hof02,Shi02}. Peacock and Robertson \cite{Pea03} showed that hydrogen acts as a shallow donor in many oxides, while deep in the silicates, SiO$_2$ and Al$_2$O$_3$. Scanlon and Watson \cite{Sca11} studied the behavior of hydrogen in p-type Cu$_2$O; they showed that hydrogen could weaken the performance of Cu$_2$O solar cell devices. It is important, however, to study the H atom influence on different properties of Cu$_2$O and M doped Cu$_2$O (M= Na, Li, Ti). As the fabrication process of solar cell devices uses different kind of gas, the knowledge of the hydrogen effect is important in choosing the gas in the annealing process.
 
Our work uses the first-principles calculations. We study the effect of hydrogen on the electronic, magnetic and optical properties of Cu$_2$O and Cu$_{2(1-x)}$M$_x$O. To find the favoured position of the hydrogen atom in the systems studied, we calculated the formation energies of different configurations proposed. In addition, we try to discuss the influence of hydrogen and different dopants on physical properties of M doped Cu$_2$O. We also predict and give an easy, low-cost, and scalable strategy to prepare the Cu$_2$O composite on heterostructure solar cells, hydrogen storage and/or production.

\section{Computational details}
We used \textit{ab initio} total-energy and molecular-dynamics program VASP (Vienna \textit{ab initio} simulation program) developed at the Fakult\"{a}t f\"ur Physik of the Universit\"{a}t Wien \cite{Kre96,Kre99}, to do all calculations. For treatment of the exchange and correlation energies, we used the generalized gradient approximation (GGA) \cite{Per69} with projector-augmented wave (PAW) \cite{Kre99,Blo94} pseudo-potentials. Brillouin zone integrals are converged with a 450 eV plane-wave cut-off, and a $2\times2\times2$ Monkhorst-Pack $\textbf{k}$-point mesh suffices to make sure that energy convergence takes place for supercell. These calculations showed a discrepancy within 10$^{-7}$ eV. We performed a relaxation with the standard conjugated gradient algorithm. We used the supercell approach to simulate the H doped Cu$_2$O or (Cu$_2$, M)O system. The optimized lattice constant of Cu$_2$O, $a = 4.26$ {\AA}, is in good agreement with the experimental value of $a = 4.27$ {\AA} \cite{Hal70,Man74} and theoretical values $a= 4.18$ and 4.20 {\AA} \cite{Kor11,Las03}.

In these calculations, we positioned the M atoms at the cation sites (Cu). We performed the H defect formation energy calculations on $3\times3\times3$ supercell, containing 162 atoms, using the calculated lattice constant with the defect at different positions in the cell. The supercell size is necessary to allow for a detailed study of many different dopant geometric structures. For calculations of the electronic properties of H doped (Cu$_2$, M)O, we suppose that M atom leads to the formation of $3\times3\times3$ supercell with chemical composition Cu$_{2(107)}$MO$_{54}$, with one atom of hydrogen at the interstitial site. All calculations were spin polarized.
We can determine optical properties using the complex dielectric function $\varepsilon(\omega)=\varepsilon_1(\omega)+\ri \varepsilon_2(\omega)$ \cite{Zhe12}. The imaginary part of the dielectric function $\varepsilon_2(\omega)$ was calculated from the momentum matrix elements between the occupied and unoccupied wave functions \cite{Res03}:
\begin{equation}\label{eq1}
  \varepsilon_2(\hbar\omega)=\frac{2e^2\piup}{\Omega\varepsilon_0}\sum_{\mathbf{k},V,C}\left|\left\langle\Psi_\mathbf{k}^C|\widehat{\mathbf{u}}\cdot\mathbf{r}|\Psi_\mathbf{k}^V\right\rangle\right|^2\delta\big(E_\mathbf{k}^C-E_\mathbf{k}^V-\hbar\omega\big).
\end{equation}
Here, $\Omega, C, V, \mathbf{k}, \omega, \widehat{\mathbf{u}}$ are the volume of cell, conduction band, valence band, $\mathbf{k}$ point, photon frequency, external field vector, respectively, while $\Psi_\mathbf{k}^C$ and $\Psi_\mathbf{k}^V$  represent the eigenstates. We can obtain the real part of the dielectric function, $\varepsilon_1$ from $\varepsilon_2$ by the Kramer--Kronig relationship \cite{Mel77}. The corresponding absorption coefficient was obtained using the following equation:
\begin{equation}\label{eq2}
  \alpha(\omega)=\sqrt{2}\Big(\frac{\omega}{c}\Big)\Big[\sqrt{\varepsilon_1^2(\omega)-\varepsilon_2^2(\omega)}-\varepsilon_1(\omega)\Big]^{\frac{1}{2}}.
\end{equation}
\section{Results and discussion}
\subsection{Hydrogen in Cu$_2$O}
\begin{figure}[!t]
  \centering
  \includegraphics[scale=0.28]{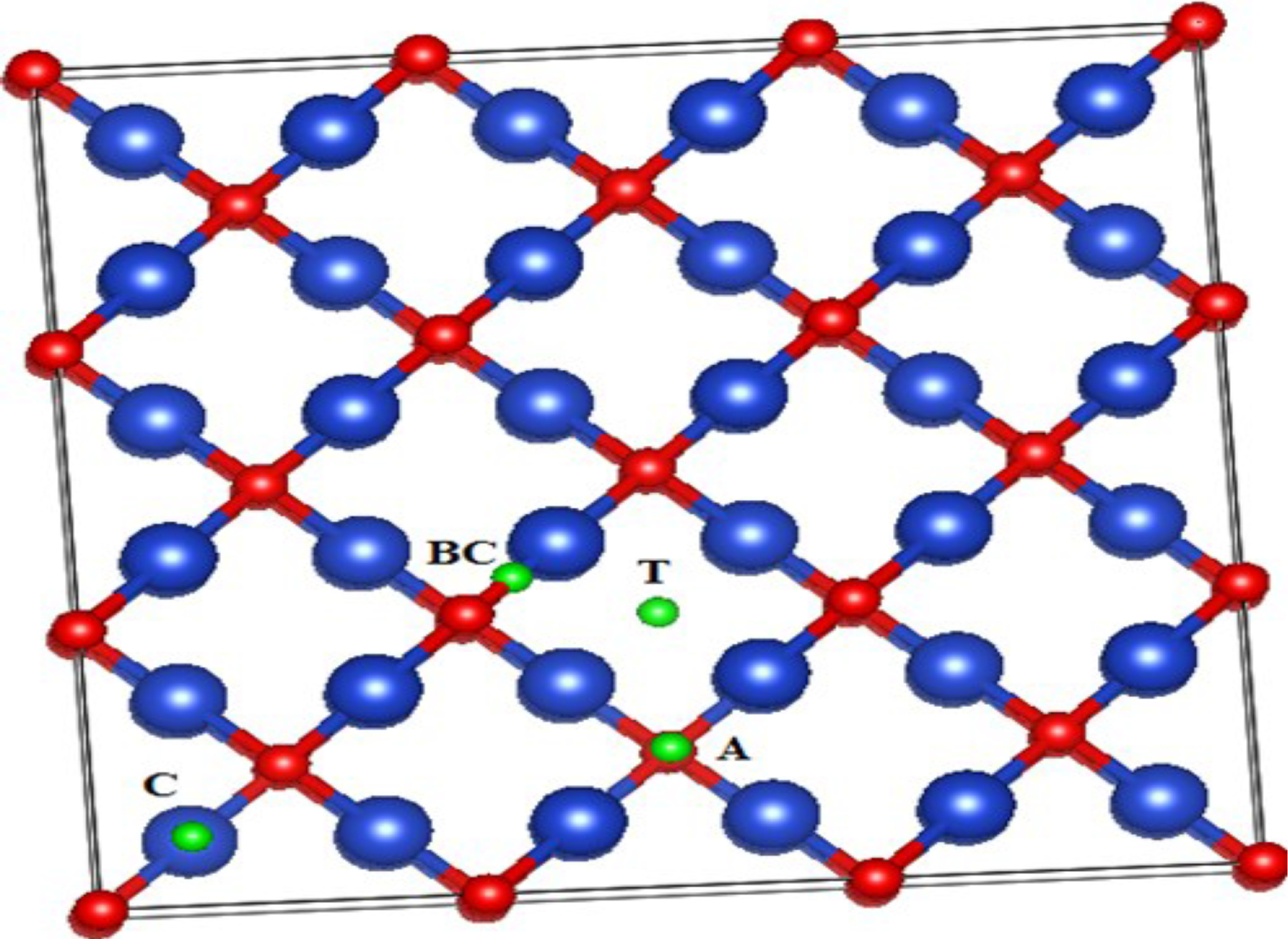}
  \caption{(Colour online) Location of H impurity in Cu$_2$O lattice. Cu atoms in blue color, O atoms in red and H atoms are in green.}\label{Fig1}
\end{figure}

In Cu$_2$O lattice we can attribute different substitutional and interstitial positions to hydrogen impurity. Figure~\ref{Fig1} illustrates all positions where copper, oxygen and hydrogen atoms are, respectively, represented by blue, red and green spheres. In this figure, we notice four possible sites for the hydrogen impurity, T~(Tetrahedral site), BC (Bond Center site), C (Cation substitutional site) and A (Anion substitutional site).
%%%%%%%%%%%%%%%%%%%%%%%%%%%%%%%%%%%%%%
%%%%%%%%%%%%%%%%%%%%%%%%%%%%%%%%%%%%
To find the preferred site of H in Cu$_2$O, we calculated the formation energies for all hydrogen positions. We show the obtained results in table~\ref{tab1}.
The references \cite{Van89,Neu99} give the formation energy for hydrogen insertion in Cu$_2$O as follows:
\begin{equation}\label{eq3}
  E_\text{f}=E_{\text{Cu}_2\text{O}+\text{H}}-E_{\text{Cu}_2\text{O}}-E_\text{tot}^\text{H}+qE_\text{F}\,.
\end{equation}
In case of substitution, the formation energy associated is:
\begin{equation}\label{eq4}
  E_\text{f}=E_{\text{Cu}_2\text{O}+\text{H}}-E_{\text{Cu}_2\text{O}}-E_\text{tot}^\text{H}+E_\text{tot}^\text{Cu}+qE_\text{F}\,,
\end{equation}
\begin{equation}\label{eq5}
E_\text{f}=E_{\text{Cu}_2\text{O}+\text{H}}-E_{\text{Cu}_2\text{O}}-E_\text{tot}^\text{H}+E_\text{tot}^\text{O}+qE_\text{F}\,.
\end{equation}
%%%%%%%%%%%%%%%%%%%%%%%%%%%%%%%
$E_{\text{Cu}_2\text{O}+\text{H}}$ is the total energy of supercell with one hydrogen atom,  $E_{\text{Cu}_2\text{O}}$ is the total energy of supercell without defects $E_\text{tot}^\text{H}$ , $E_\text{tot}^\text{Cu}$ and $E_\text{tot}^\text{O}$ are the total energies of an isolated hydrogen, cuprous and oxygen atoms, respectively, in the ground state. The last term in the 
formation energy ($q$) accounts because H+ donates an electron $(q=-1)$, and H- 
accepts an electron $(q=+1)$. $E_\text{F}$ is the Fermi level energy. If the formation energy is positive $(E_\text{f}>0)$, the implantation of the hydrogen atom into the Cu$_2$O lattice is energetically unfavorable.

\begin{figure}[!t]
  \centering
  \includegraphics[scale=0.4]{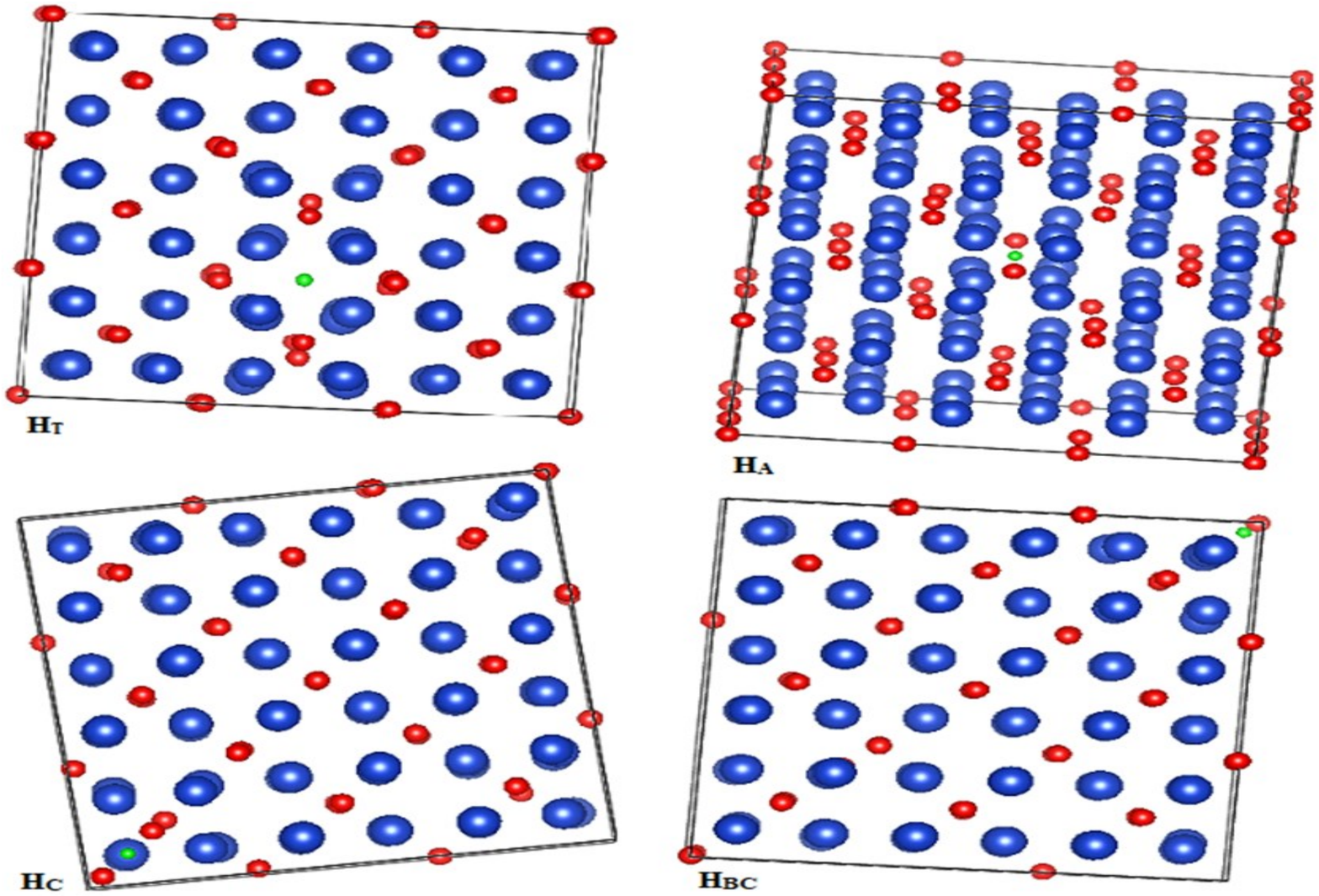}
  \caption{(Colour online) H impurity in Cu$_2$O lattice after relaxation. Cu atoms in blue color, O atoms in red and H atoms are in green.}\label{Fig2}
\end{figure}

One can see from table~\ref{tab1} that the lowest formation energy is that of the T site with $-0.77$ eV. These values show that hydrogen prefers to localize in the tetrahedral site in Cu$_2$O lattice, which confirms earlier results \cite{Sca11,Cox06}. Figure~\ref{Fig2} shows different configurations of hydrogen in Cu$_2$O relaxed lattice.  We can see that with insertion in the tetrahedral site, anionic site and cationic site, the hydrogen position does not change, while with insertion in Cu-O bond center, the hydrogen atom takes another position in the Cu$_2$O lattice.
%%%%%%%%%%%%%%%%%%%%%%%%%%%%%%%%%%%%%%%%%
\begin{table}[!b]
	\vspace{-4mm}
	\caption{Formation energies for different sites of hydrogen in Cu$_2$O lattice.}
	\label{tab1}
	\begin{center}
		\begin{tabular}{|c|c|c|c|c|}
			\hline
			Sites &	H$_\text{T}$ & H$_\text{BC}$ &	H$_\text{A}$ &	H$_\text{C}$\\
			\hline
			$E_\text{f}$(eV) & -- 0.77 & 5.64 & 0.15 & 0.41 \\
			\hline
		\end{tabular}
	\end{center}
\end{table}
%%%%%%%%%%%%%%%%%%%%%%%%%%%%%%%%%%%%%%
%%%%%%%%%%%%%%%%%%%%%%%%%%%%%%%%%%%%%%%%
%%%%%%%%%%%%%%%%%%%%%%%%%%%%%%%%%%%%%%
%%%%%%%%%%%%%%%%%%%%%%%%%%%%%%%%%%%%
%
%%%%%%%%%%%%%%%%%%%%%%%%%%%%%%%%

\begin{figure}[!b]
       \centerline{\includegraphics[scale=0.3]{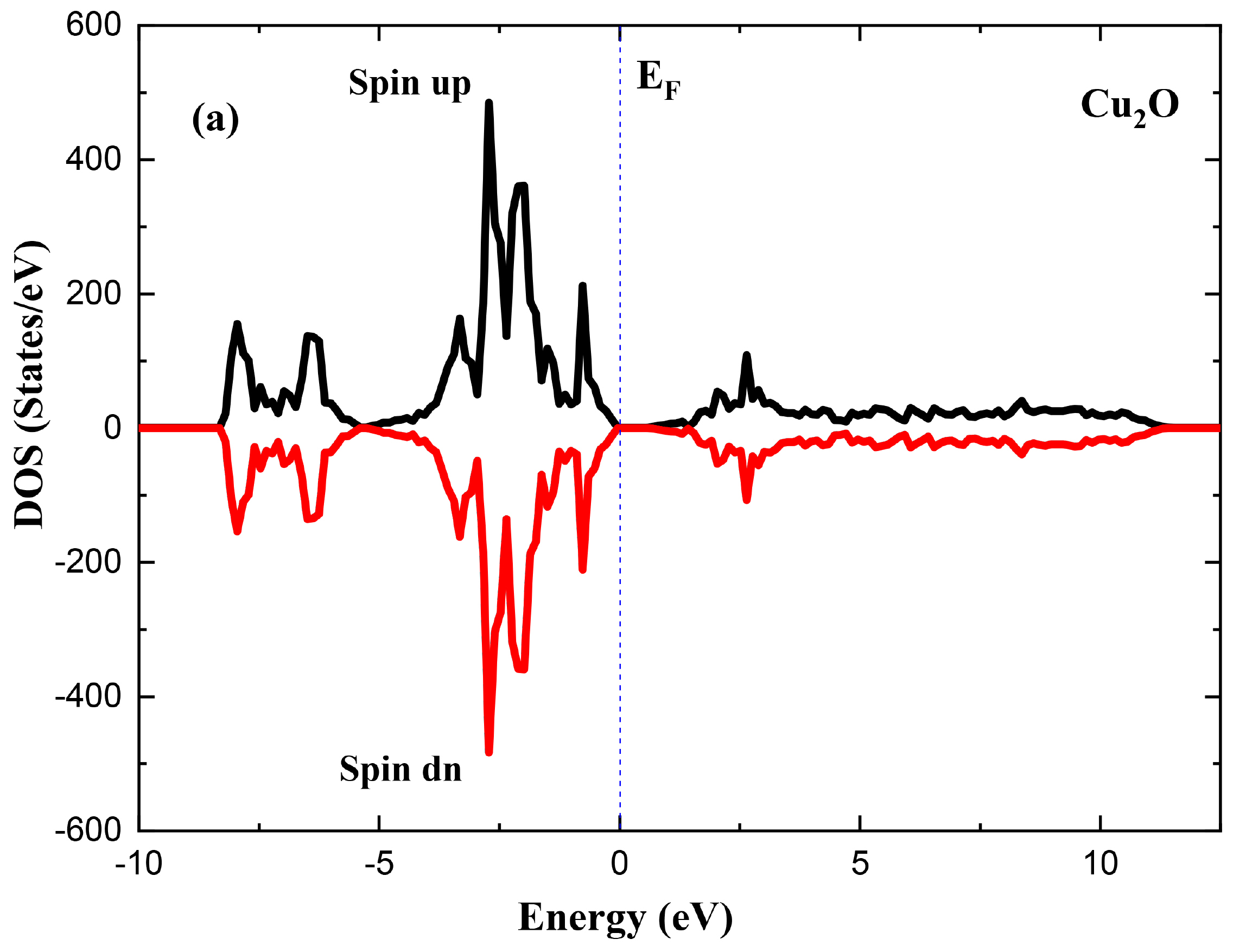}
    \includegraphics[scale=0.3]{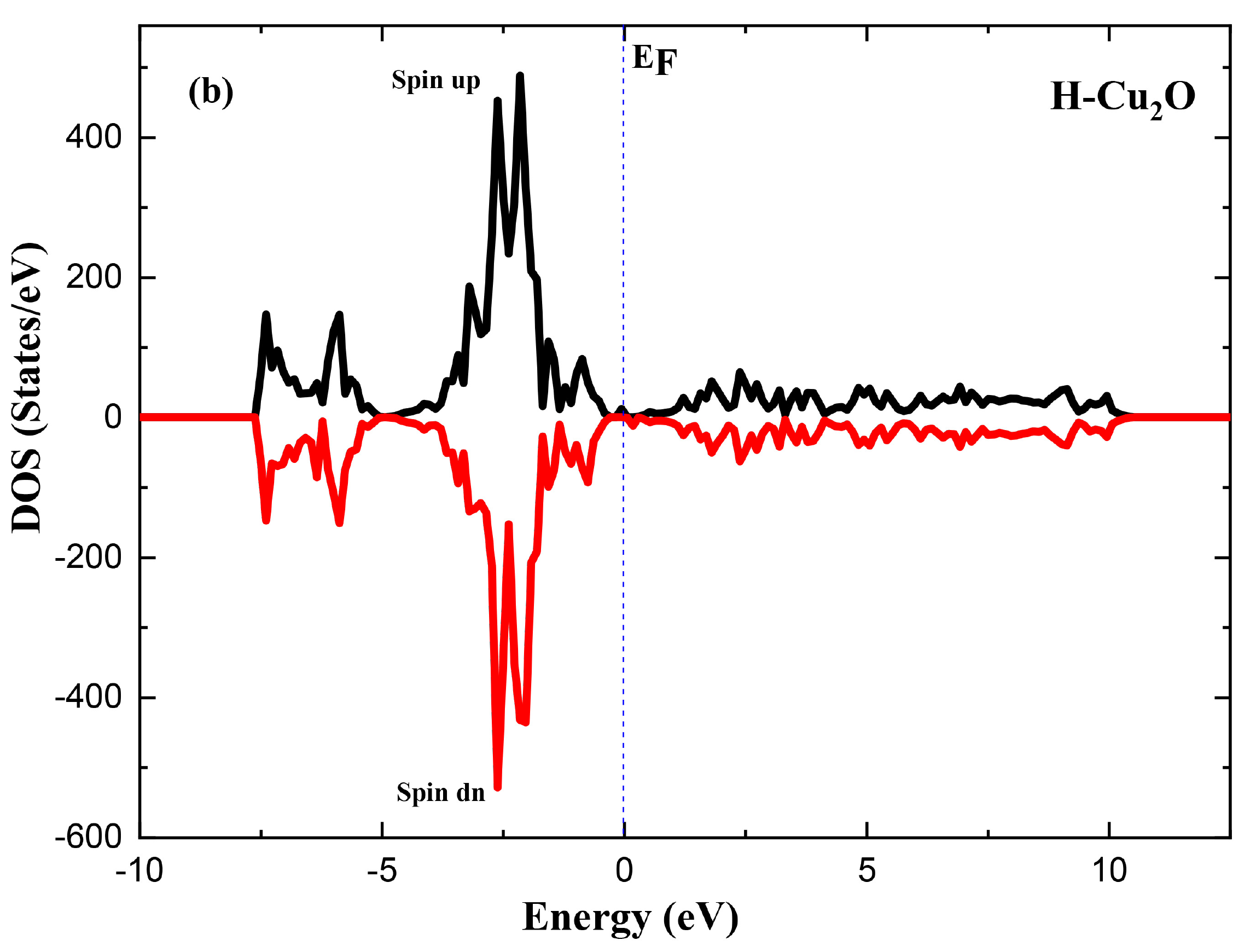}}
    \caption{(Colour online) Total density of states TDOS: (a) Pure Cu$_2$O, (b) H doped Cu$_2$O.}
    \label{fig3}
\end{figure}

We plotted the total densities of state (TDOS) of pure Cu$_2$O and a single H-doped Cu$_2$O in figure~\ref{fig3}. The TDOS figures show that hydrogen causes a change in the position of the Fermi level from near the valence band towards conduction band, leading the system to n-type semiconductor characteristic; this result agrees well with earlier works \cite{Sca11,Mat12}. In fact, this behavior of hydrogen was found in other p-type oxide materials such as SnO$_2$, TiO$_2$ and ZnO \cite{Mat12,Mol54,Tho56,Lan57}. The calculated band gap of Cu$_2$O, $E_\text{g}$=0.55 eV, is smaller than the experimental value (2.17 eV) \cite{Kit86} according to the DFT-GGA calculations. The result agrees well with the earlier theoretical values reported in the literature (0.6 eV \cite{Mar86,Mar03}, 0.64 eV \cite{Soo06} and 0.7 eV \cite{Isl09}). For hydrogen doped Cu$_2$O, we notice a narrow energy gap of 0.09 eV of the majority-spin electrons, while for the minority-spin electrons, the band gap is 0.23 eV. The hydrogen in Cu$_2$O decreases the band gap value of Cu$_2$O.

To improve the physical properties of Cu$_2$O for voltaic applications, we have tried three different dopants: Sodium (Na), Lithium (Li) and Titanium (Ti). We investigated the properties of M doped Cu$_2$O, M= Na, Li and Ti,  by substituting one M atom for one Cu atom in the $3\times3\times3$ supercell Cu$_{(108)}$O$_{54}$.

In Cu$_{2(1-x)}$M$_x$O structure, we consider two configurations where interstitial hydrogen atom is at the tetrahedral site. In the first one, a short distance (S$_\text{D}$) separates M and H while in the second one, a long distance (L$_\text{D}$) separates M and H atoms.

\subsection{Na and H doped Cu$_2$O}
In the first case configuration (S$_\text{D}$), the distance between Na and H was 1.78 {\AA} while after relaxation this distance becomes 2.38 {\AA}. This suggests that H atom relaxes to move away from the Na atom. For the second configuration (L$_\text{D}$) and before relaxation, the distance between Na and H was 6.98 {\AA}. After relaxation, this distance hardly moves to only 7.012 {\AA}.

\begin{figure}[!t]
       \centerline{\includegraphics[scale=0.3]{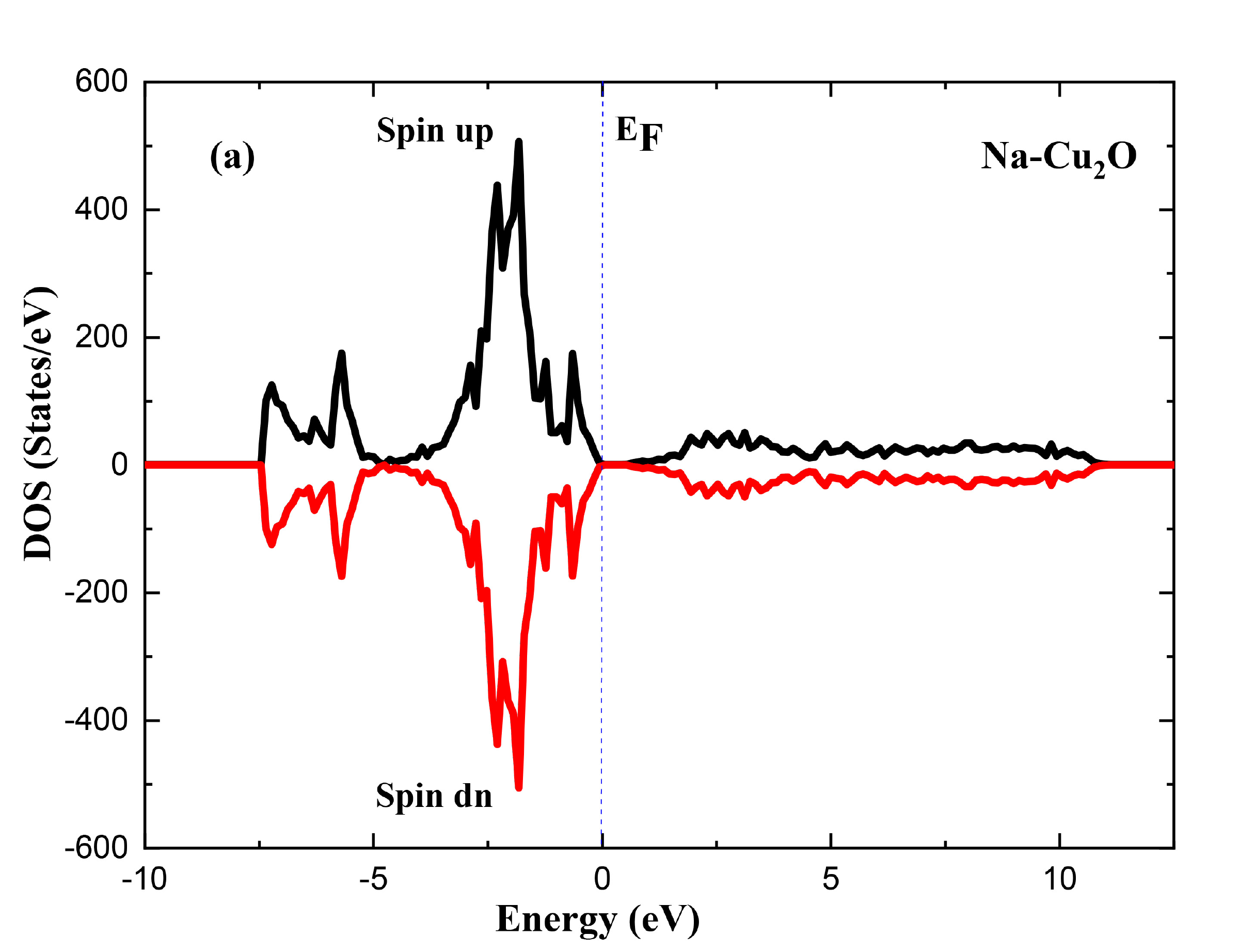}
    \includegraphics[scale=0.3]{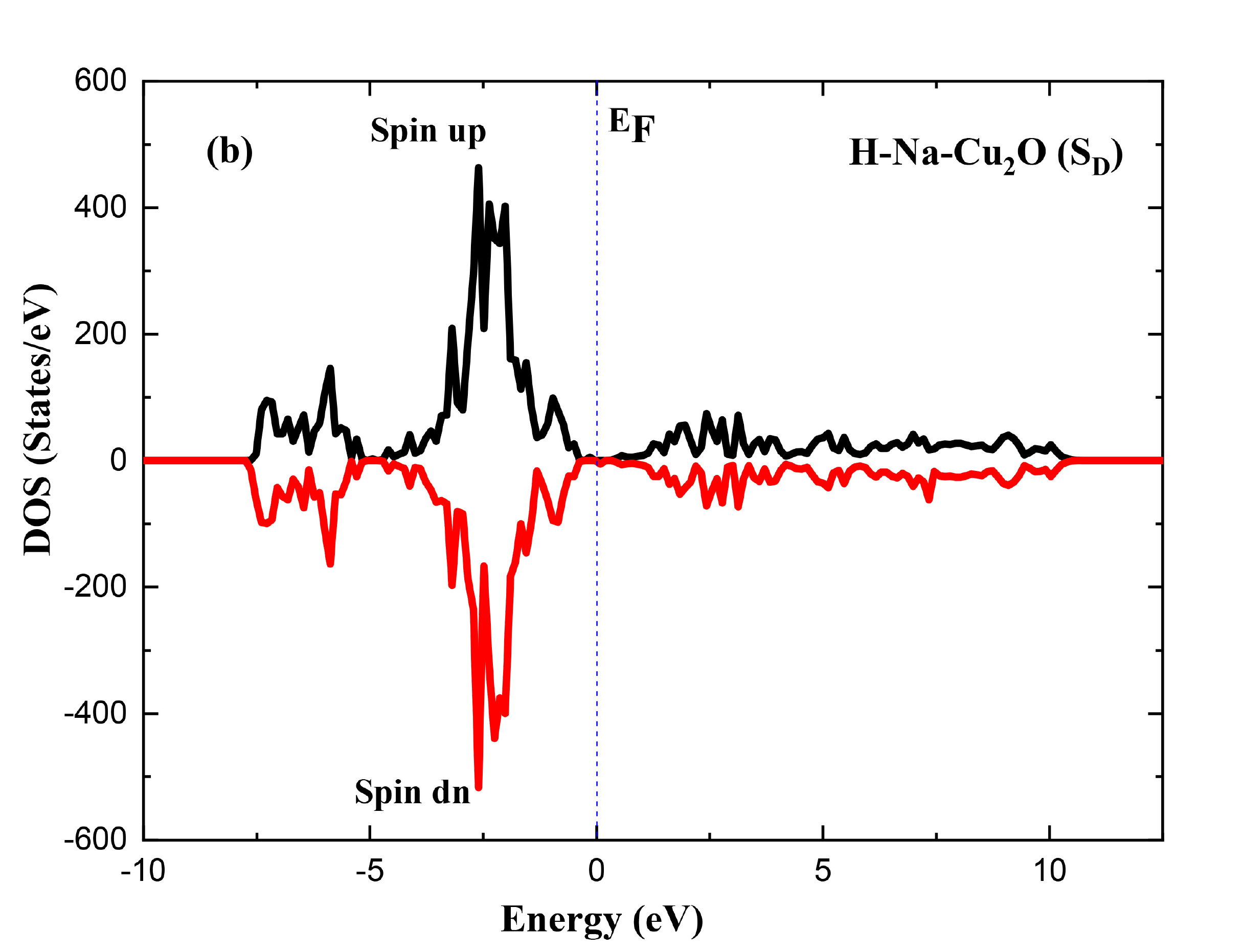}}
     \centerline{\includegraphics[scale=0.3]{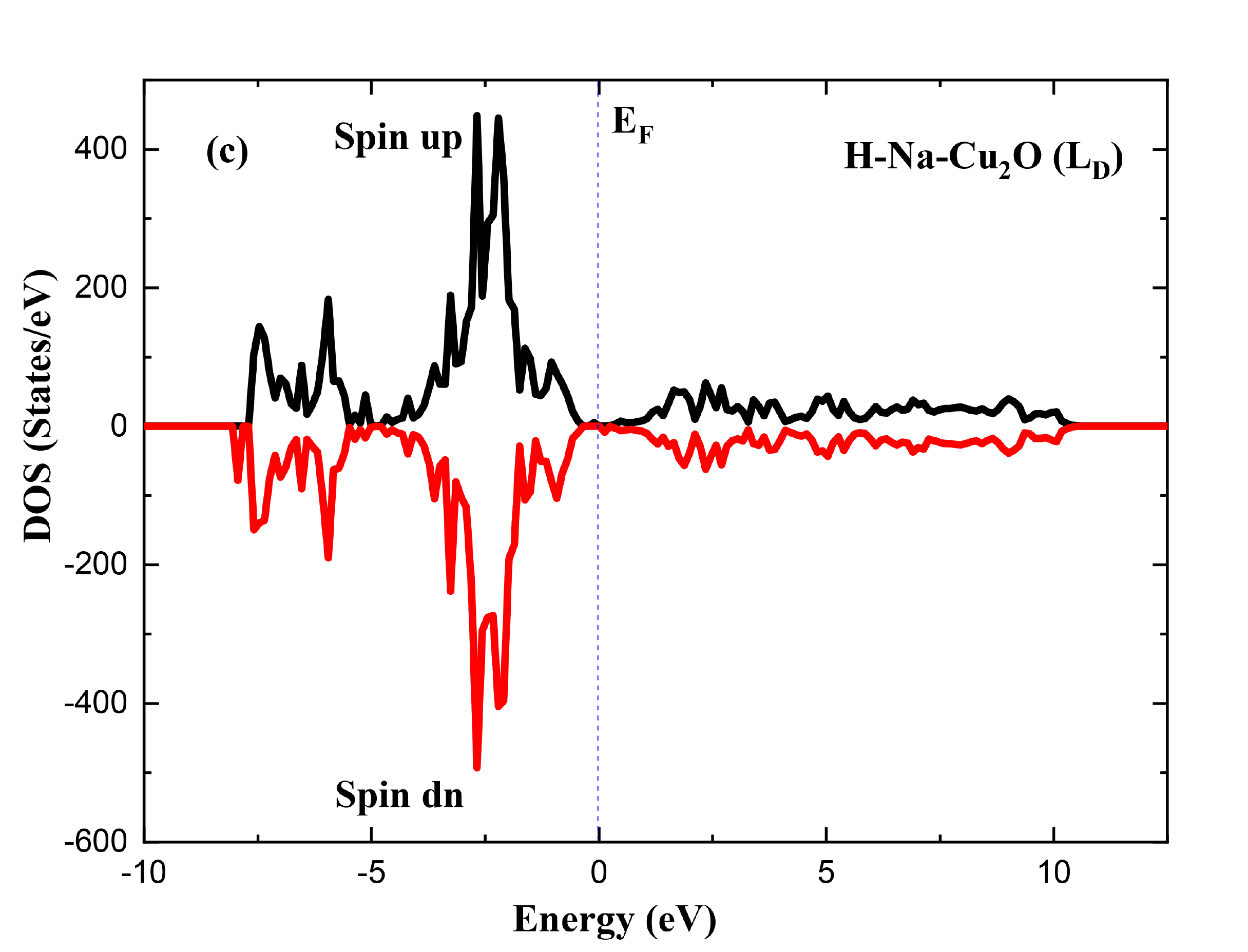}}
    \caption{(Colour online) Total density of states TDOS: (a) Na-doped Cu$_2$O, (b) H-Na-doped Cu$_2$O (S$_\text{D}$), (c) H-Na-doped Cu$_2$O (L$_\text{D}$).}
    \label{fig4}
\end{figure}

We show the total density of states of Na doped Cu$_2$O in figure~\ref{fig4}~(a). The figure indicates that the Na doping decreases the band gap of Cu$_2$O. The doping of Cu$_2$O by Na does not change the nature of his conductivity; the calculated energy of the gap is 0.35 eV. Elfadill \textsl{et al}. \cite{Nez16} showed that Na doping Cu$_2$O is a p-type characteristic, which shows that our calculations are in good agreement with experimental data. In addition, Minami \textsl{et al}. used Cu$_2$O doped Na as p-type in p-n heterojunction and/or homojunction solar cells \cite{Tad15,Tad16}.
%%%%%%%%%%%%%%%%%%%%%%%%%%%%%
%%%%%%%%%%%%%%%%%%%%%%%%%%%%%%

For hydrogen implantation in Na doped Cu$_2$O, the calculated band gap is 0.47 eV for a short distance. In the long distance case, we note two different band gap values, 0.47 eV and 0.35 eV for the majority-spin and minority-spin electron, respectively. From the DOS figures [figure~\ref{fig4}~(b), figure~\ref{fig4}~(c)], we note that hydrogen changes the position of Fermi level ($E_\text{F}$) of Na doped Cu$_2$O, $E_\text{F}$ moves from the valence band to conduction band. Thus, Na doped Cu$_2$O becomes n-type semiconductor. The distance between Na and H atoms does not influence the electronic properties of Na doped Cu$_2$O, we found the same band gap value in the two configurations.
\subsection{Li and H doped Cu$_2$O}
For the short distance case, the distance between Li and H was 1.78 {\AA}. After the relaxation, the distance becomes 1.69 {\AA} while in the long distance case, the distance between Na and H was 5.27 {\AA}. After the relaxation, the distance becomes 5.35 {\AA}.
%%%%%%%%%%%%%%%%%%%%%%%%%%%%%
\begin{figure}[!t]
       \centerline{\includegraphics[scale=0.3]{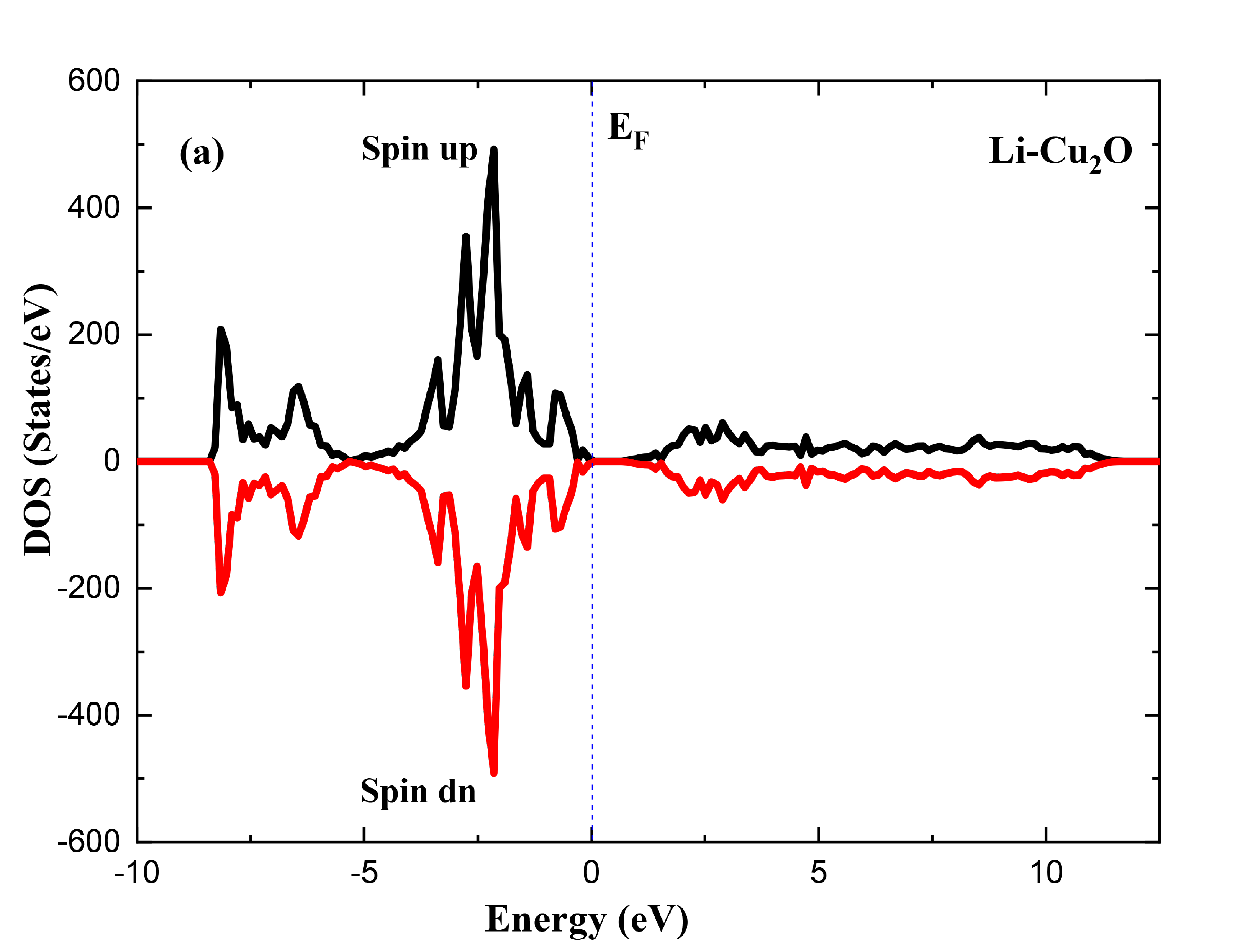}
    \includegraphics[scale=0.3]{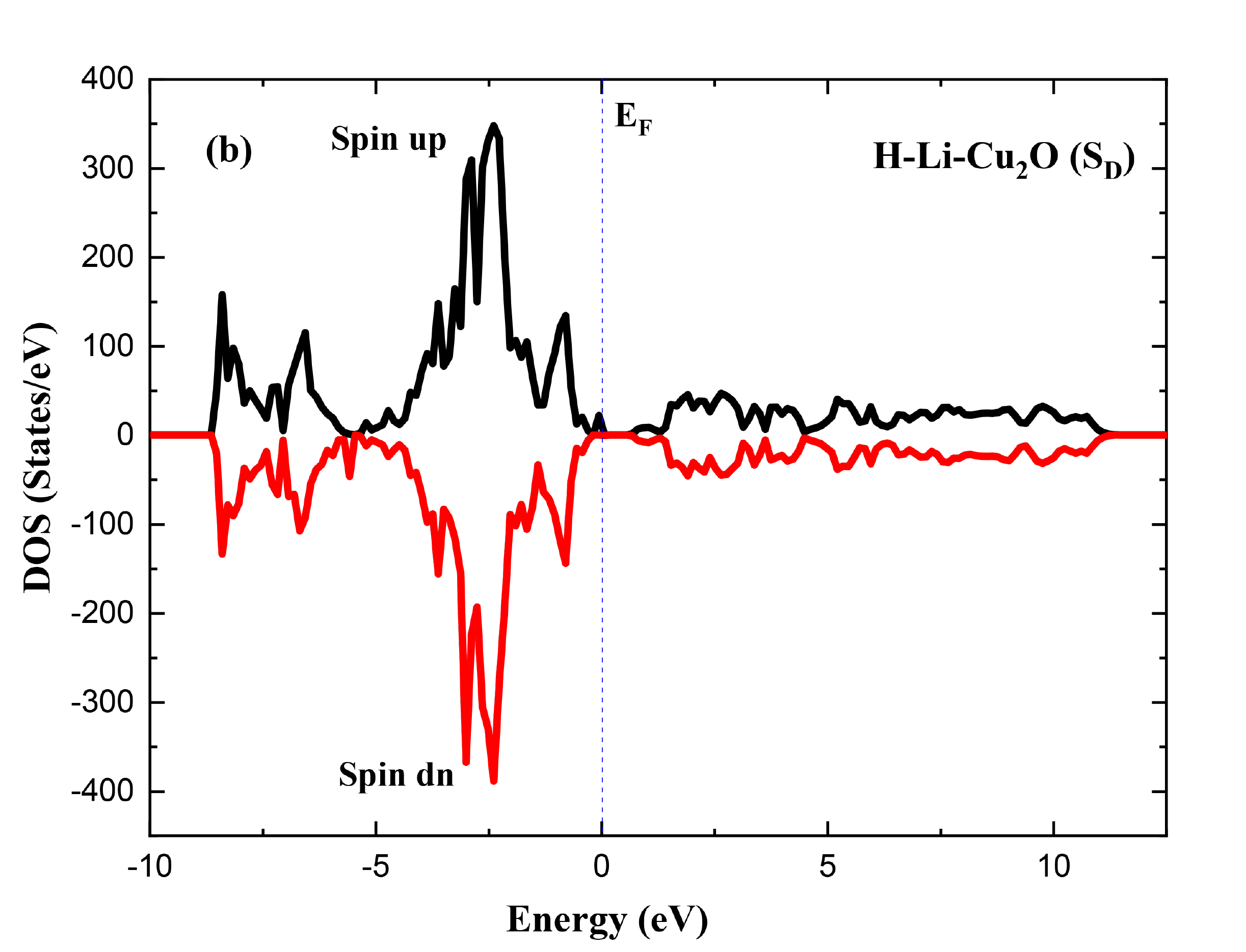}}
     \centerline{\includegraphics[scale=0.3]{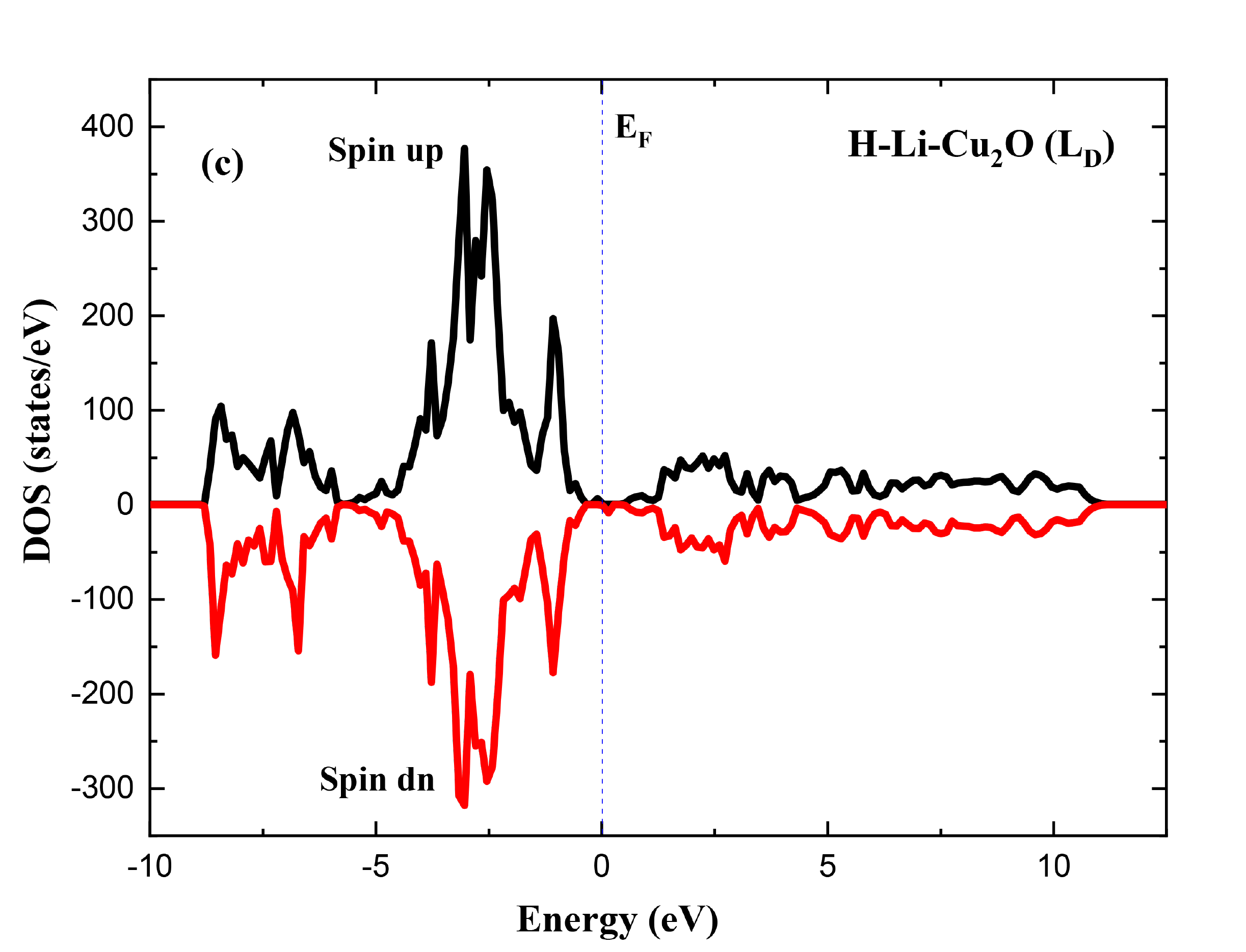}}
    \caption{(Colour online) Total density of states TDOS: (a) Li-doped Cu$_2$O, (b) H-Li-doped Cu$_2$O (S$_\text{D}$), (c) H-Li-doped Cu$_2$O (L$_\text{D}$).}
    \label{fig5}
\end{figure}
%%%%%%%%%%%%%%%%%%%%%%%%%%%%%%

Figure~\ref{fig5}~(a) shows the total densities of states for Li-doped Cu$_2$O. From Fermi level position we note that the Li doped Cu$_2$O is p-type characteristic. This result agrees well with the experimental work of Kyung-Su Cho \textsl{et al}. \cite{Cho2016} where they used Li doped Cu$_2$O as p-type in p-n heterojunction. The calculated band gap of Li-doped Cu$_2$O is 0.61 eV.

From the DOS figures [figure~\ref{fig5}~(b), figure~\ref{fig5}~(c)] of H-Li-doped, we see the shift in the Fermi level position. The hydrogen implantation in Li-doped Cu$_2$O does not change the value of the band gap but it makes the Cu$_2$O an n-type semiconductor characteristic. The distance between Li and H does not influence the band gap value of Li-doped Cu$_2$O.
\subsection{Ti and H doped Cu$_2$O}
For the first configuration (S$_\text{D}$), the distance between Ti and H is 1.8 {\AA}. After the relaxation, the distance becomes 1.88 {\AA}. In the second case configuration (L$_\text{D}$), the distance between Ti and H was 7.92 {\AA}. After the relaxation, the distance becomes 7.9 {\AA}.
%%%%%%%%%%%%%%%%%%%%%%%%%%%%%
\begin{figure}[!t]
       \centerline{\includegraphics[scale=0.3]{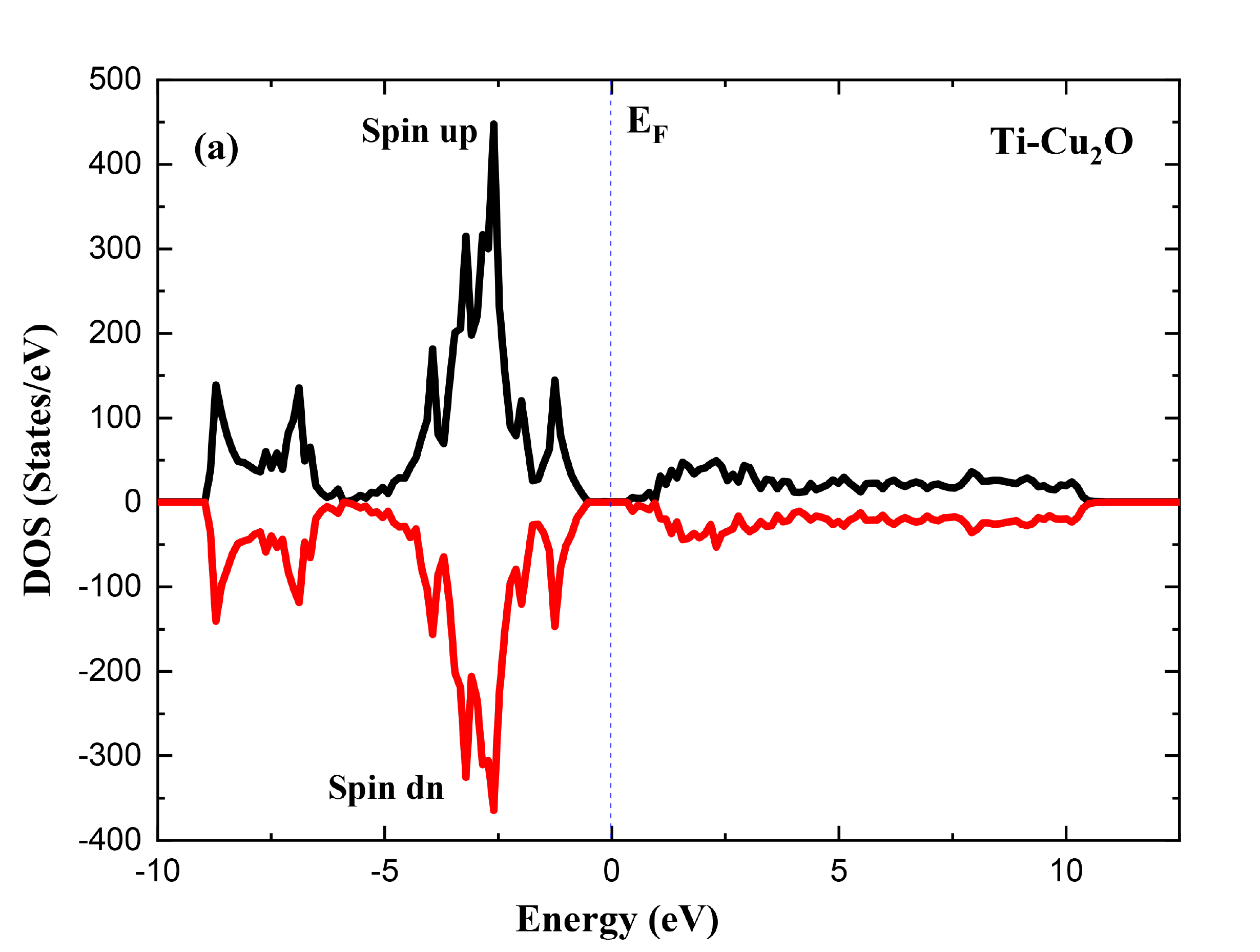}
    \includegraphics[scale=0.3]{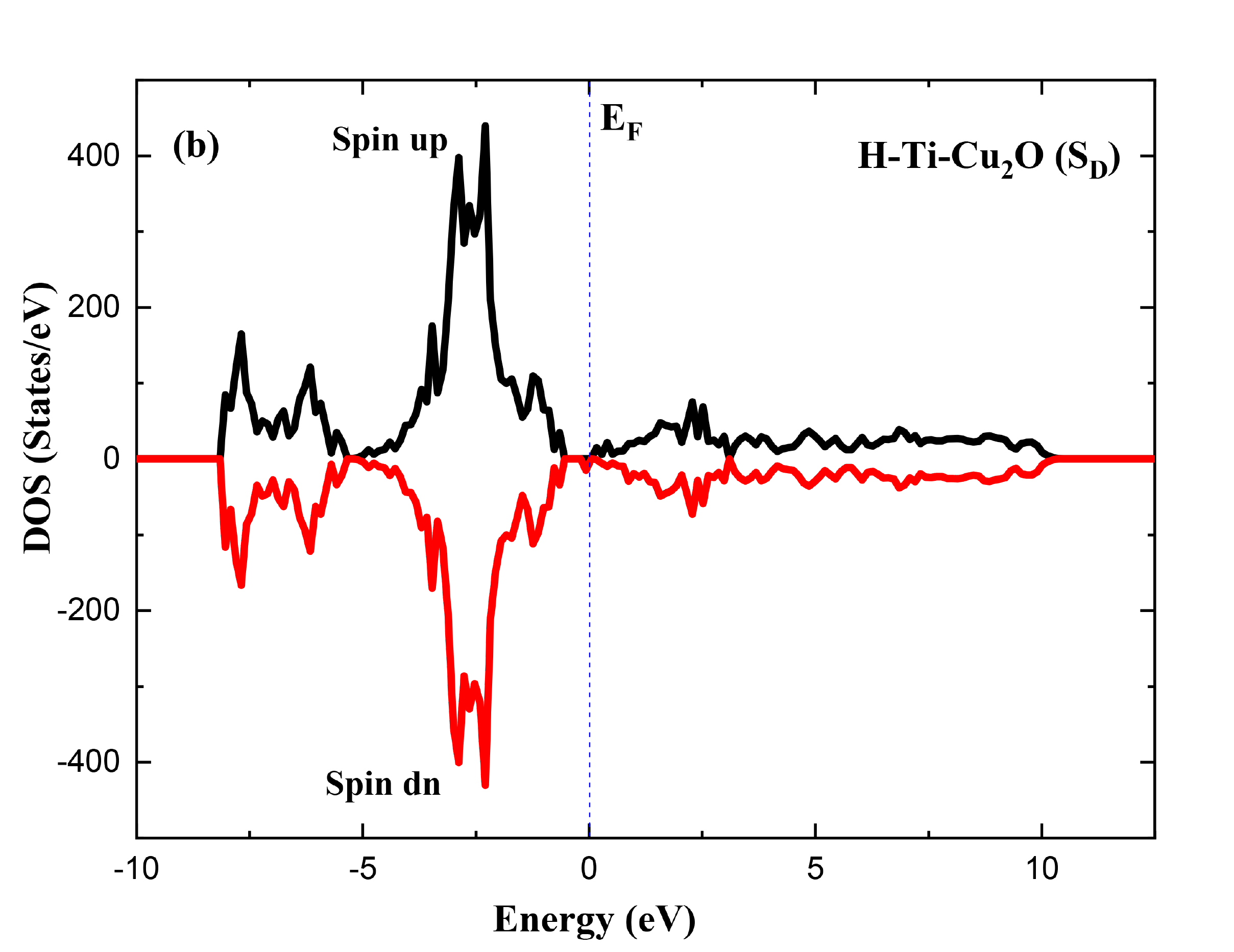}}
     \centerline{\includegraphics[scale=0.3]{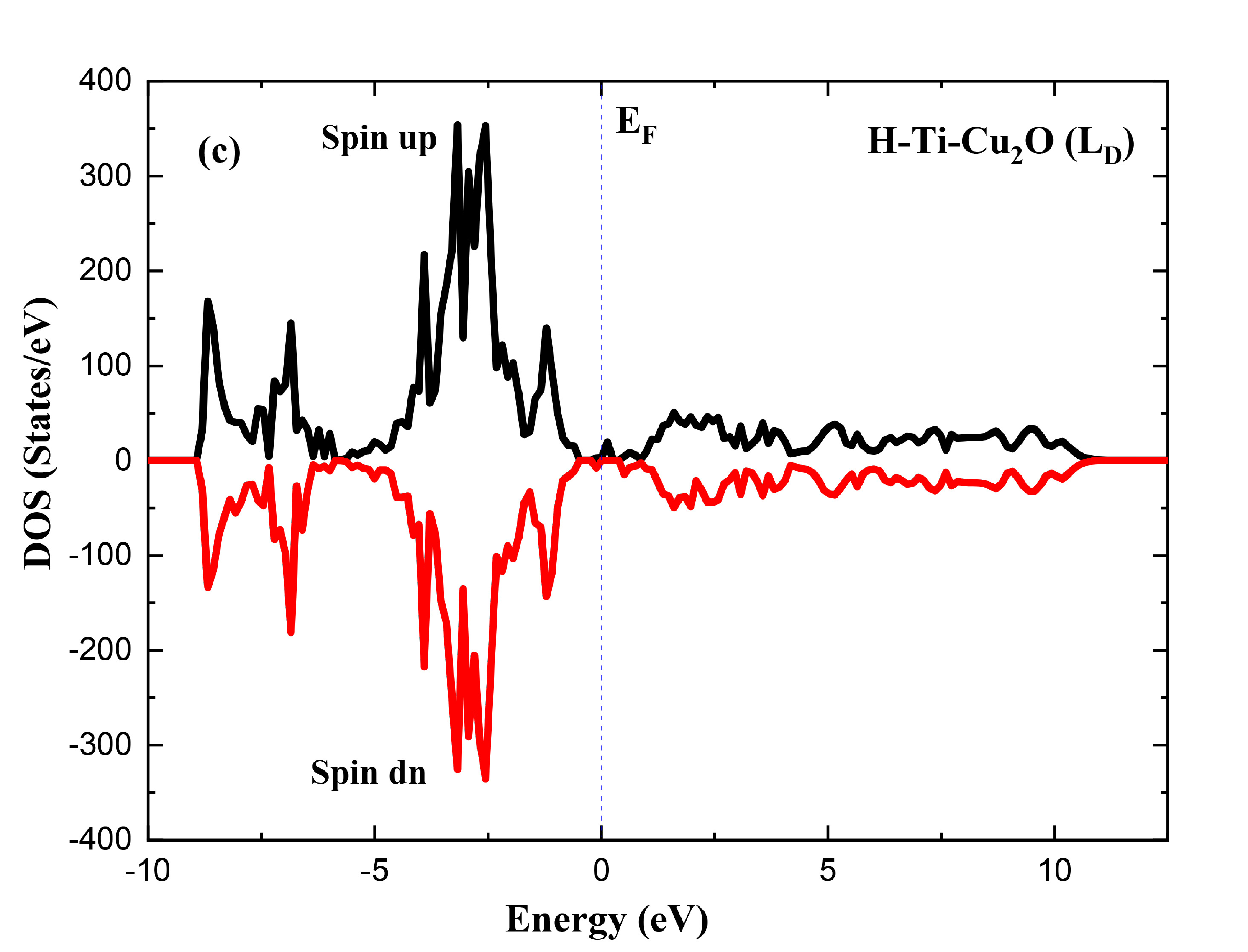}}
    \caption{(Colour online) Total density of states TDOS: (a) Ti-doped Cu$_2$O, (b) H-Ti-doped Cu$_2$O (S$_\text{D}$), (c) H-Ti-doped Cu$_2$O (L$_\text{D}$).}
    \label{fig6}
\end{figure}
%%%%%%%%%%%%%%%%%%%%%%%%%%%%%%

Figure~\ref{fig6} shows the total densities of states for Ti-doped Cu$_2$O and H-Ti-doped Cu$_2$O. The calculated band gap of Ti-doped Cu$_2$O is 0.73 eV and 0.86 eV for the majority-spin and minority-spin electron, respectively. These values are larger than that of the pure Cu$_2$O (0.55 eV). From the DOS figures, we note that Ti doping changes the position of Fermi level from near the valence band towards the conduction band, leading the system to n-type semiconductor characteristic. In the first case configuration, the hydrogen implantation can decrease the band gap value of Ti doped Cu$_2$O to 0.59 eV, while in the second case configuration, hydrogen does not change the band gap value. In the two configurations, it does not change the conductivity type of Ti doped Cu$_2$O.
%%%%%%%%%%%%%%%%%%%%%%%%%%%%%%%%%%%%%%%%%
\begin{table}[!t]
\caption{Calculated band gap $E_\text{g}$(eV) and lattice parameter $a$({\AA}).}
\label{tab2}
\begin{center}
\begin{tabular}{|c|c|c|}
\hline
  &	$E_\text{g}$(eV) & $a$({\AA}) \\
 \hline
Cu$_2$O & 0.55 & 4.26 \\
 \hline
 H-Cu$_2$O (spin up) &	0.46 & 4.28 \\
 \hline
 H-Cu$_2$O (spin dn) & 0.23	 & -- \\
 \hline
 Na-Cu$_2$O & 0.35	 & 4.26 \\
 \hline
H-Na-Cu$_2$O (S$_\text{D}$) &	0.47 & 4.27 \\
 \hline
H-Na-Cu$_2$O (L$_\text{D}$) (spin up) & 0.47 & 4.27 \\
 \hline
H-Na-Cu$_2$O (L$_\text{D}$) (spin dn) & 0.35 & -- \\
 \hline
Li-Cu$_2$O & 0.61 & 4.12 \\
 \hline
H-Li-Cu$_2$O (S$_\text{D}$) &	0.61 & 4.11 \\
 \hline
H-Li-Cu$_2$O (L$_\text{D}$) & 0.61 & 4.11 \\
 \hline
Ti-Cu$_2$O (spin up) &	0.73 & 4.11 \\
 \hline
Ti-Cu$_2$O (spin dn) & 0.86 & -- \\
 \hline
H-Ti-Cu$_2$O (S$_\text{D}$) &	0.59 & 4.24 \\
 \hline
H-Ti-Cu$_2$O(L$_\text{D}$) & 0.86 & 4.12 \\
 \hline
\end{tabular}
\end{center}
\end{table}
%%%%%%%%%%%%%%%%%%%%%%%%%%%%%%%%%%%%%%

Table~\ref{tab2} shows the calculated band energies ($E_\text{g}$) and lattice parameters (a) for Cu$_2$O and M doped Cu$_2$O with or without hydrogen. From this table, we can see that the band gap value of Na doping Cu$_2$O makes it the best of the three as a p-type dopant in Cu$_2$O. Compared with the value of pure Cu$_2$O, the lattice parameter of Na doped Cu$_2$O does not change while Li and Ti doping should compress the lattice of Cu$_2$O. For all cases, hydrogen does not change the lattice of systems except with hydrogen implantation in the first configuration (S$_\text{D}$) in Ti doped Cu$_2$O. In this configuration, hydrogen should expand the lattice of this system from 4.11 to 4.24 {\AA}.

From the densities of states figures, we can see that hydrogen creates new states in the band gap in all systems. Na and Li act as acceptor impurities in Cu$_2$O and hydrogen should passivate these acceptors. Ti is a source of n-type conductivity and hydrogen could not passivate this donor in Cu$_2$O.

In the cases of H doped Cu$_2$O, Ti doped Cu$_2$O and H-Na-co-doped Cu$_2$O for the second configuration~(L$_\text{D}$), the diference in band gap value of majority-spin and minority-spin electrons shows anisotropy of the gap which determines the elastic distribution between the hole states and electrons.

From these results, it seems that hydrogen can passivate the acceptors (Na, Li) or donors (Ti) in Cu$_2$O. Similar results have been found in experimental works; Lu \textsl{et al}. \cite{Yan05} revealed that a treatment of N-doped Cu$_2$O films in hydrogen plasma could passivate the carrier traps. On the other hand, Ishizuka~\textsl{et~al}. \cite{Ish02,Ish03} show that a treatment of N-doped Cu$_2$O films in hydrogen plasma leads to an increase of conductivity.

\subsection{Magnetic and optical properties}
We list the values of the total and local magnetic moments in table~\ref{tab3}. With hydrogen, we have the values of the total magnetic moments approximately equal to 1 $\mu_\text{B}$ except with Ti-doped Cu$_2$O, where it is almost zero. In all systems, the Cu moments dominate the values of local magnetic moments with a maximum value of 0.16 $\mu_\text{B}$ excluding the H-Ti-codoped Cu$_2$O system where local magnetic moments values are negative.
%%%%%%%%%%%%%%%%%%%%%%%%%%%%%%%%%%%%%%%%%
\begin{table}[!b]
\caption{Magnetic moments (in unit of $\mu_\text{B}$) of the supercell ($m_\text{tot}$), H atom ($m_\text{H}$), Na ($m_\text{Na}$), Li atom ($m_\text{Li}$), Ti atoms ($m_\text{Ti}$), Cu atom ($m_\text{Cu}$) and O atom ($m_\text{O}$).}
\label{tab3}
\vspace{2ex}
\begin{center}
\begin{tabular}{|c|c|c|c|c|c|c|c|}
\hline
  & $m_\text{tot}$($\mu_\text{B}$) & $m_\text{H}$($\mu_\text{B}$)& $m_\text{Na}$($\mu_\text{B}$)& $m_\text{Li}$($\mu_\text{B}$)& $m_\text{Ti}$($\mu_\text{B}$)& $m_\text{Cu}$($\mu_\text{B}$)& $m_\text{O}$($\mu_\text{B}$) \\
 \hline
 Cu$_2$O & 0.0 & -- & -- & -- & -- & 0.0 & 0.0\\
 \hline
H-Cu$_2$O  & 0.9999 & 0.056 & -- & -- & -- & 0.13 & 0.05 \\
 \hline
 Na-Cu$_2$O& 0.0 & -- & 0.0 & -- & -- & 0.0 & 0.0 \\
 \hline
 H-Na-Cu$_2$O (S$_\text{D}$)& 1.0 & 0.056 & 0.004 & -- & -- & 0.16 & 0.06\\
 \hline
H-Na-Cu$_2$O (L$_\text{D}$)& 0.9999 & 0.058 & 0.0 & -- & -- & 0.12 & 0.05\\
 \hline
 Li-Cu$_2$O& 0.0 & -- & -- & 0.0 & -- & 0.0 & 0.0\\
 \hline
 H-Li-Cu$_2$O (S$_\text{D}$)& 1.0 & 0.06 & -- & 0.013 & -- & 0.15 & 0.06\\
 \hline
 H-Li-Cu$_2$O (L$_\text{D}$)& 0.9955 & 0.06 & -- & 0.0 & -- & 0.12 & 0.04\\
 \hline
Ti-Cu$_2$O& 1.0 & -- & -- & -- & 0.75 & 0.024 & -- 0.012 \\
 \hline
 H-Ti-Cu$_2$O (S$_\text{D}$)& -- 0.0006 & 0.0 & -- & -- & 0.05 & 0.02 & -- 0.007 \\
 \hline
H-Ti-Cu$_2$O(L$_\text{D}$) & 0.0699 & -- 0.07 & -- & -- & 0.74 & -- 0.11 & -- 0.05 \\
 \hline
\end{tabular}
\end{center}
\end{table}
%%%%%%%%%%%%%%%%%%%%%%%%%%%%%%%%%%%%%%%%%%%%%%%%%%

As far as the principle of contamination of the samples by ferromagnetic impurities can be discounted, the common factor in all cases is crystal defects. Here, they have been established, but it is plausible that they are point defects-atomic vacancies or interstitials. Examples of $d^0$ ferromagnetism have been reported in the literature \cite{Coe05}.

The absorption coefficient $\alpha(\omega)$ is important to estimate the optical properties of Cu$_2$O and H-M-codoped Cu$_2$O (M= Na, Li, Ti) used in photovoltaic cells or as optoelectronic devices. A material with a low absorption coefficient means more transmitted light. Cuprous oxide Cu$_2$O is a promising p-type TCO, but its relatively low band gap (2.1 eV) limits its optical transmittance in the visible spectrum. In this paper, we aim at increasing this value.

\begin{figure}[!t]
       \centerline{\includegraphics[scale=0.55]{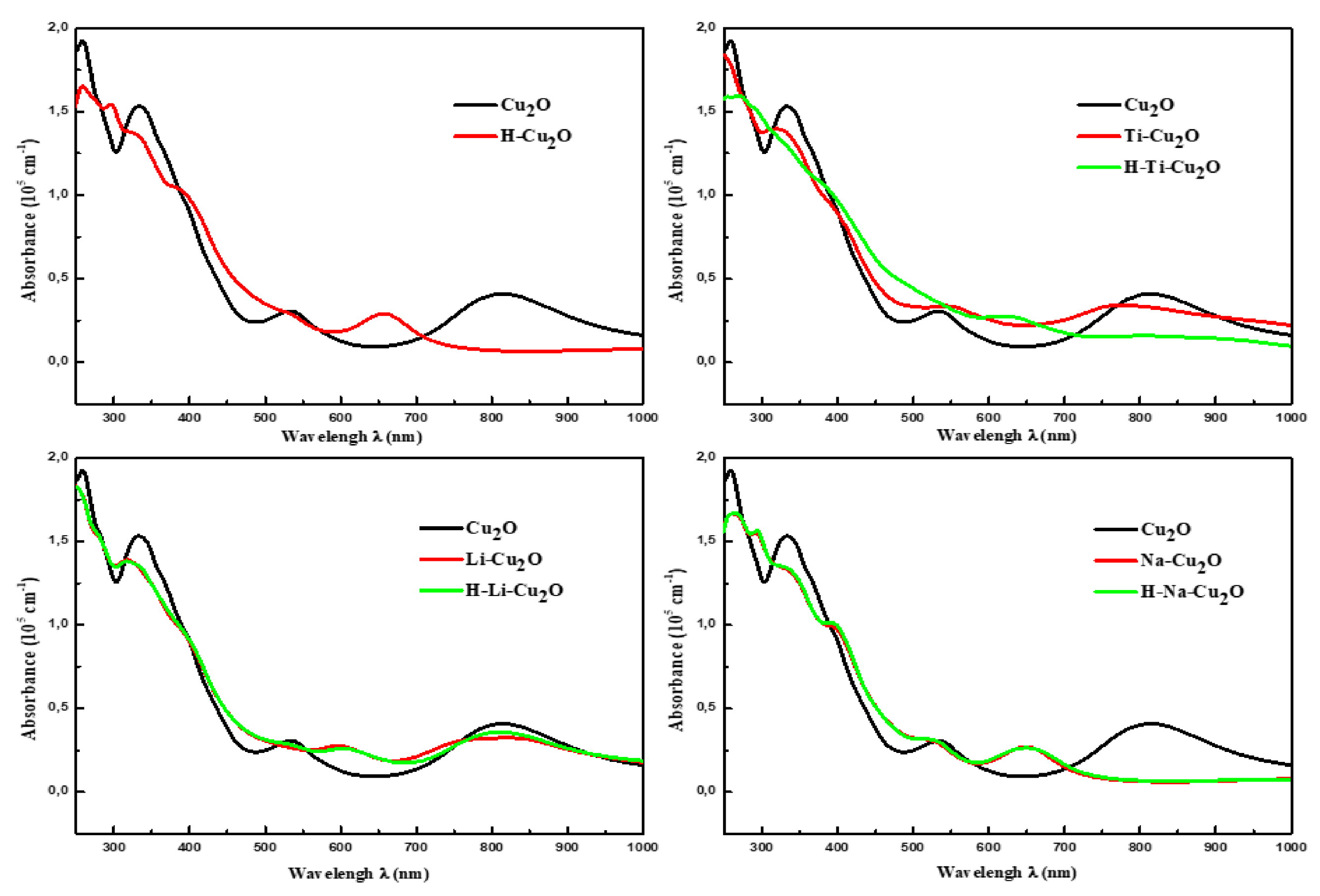}}
        \caption{(Colour online) The variation of the absorption coefficient of H inserted in M doped Cu$_2$O compared with pure one.}
    \label{fig7}
\end{figure}

Figure~\ref{fig7} presents the absorption coefficient (of the order of 104 cm$^{-1}$) versus the wavelength (nm) of all configurations compared to the pure compound Cu$_2$O in the visible light region (400--900 nm) and UV region (250--400 nm). The absorbance of pure Cu$_2$O decreases from 1.92$\times10^5$ cm$^{-1}$ to 0.89$\times10^5$~cm$^{-1}$ in (250--400 nm) and in the visible region continues to decrease to 0.086$\times10^5$ at 650 nm but we observed two peaks at 530 (0.287$\times10^5$ cm$^{-1}$) and 800 nm (0.428$\times10^5$ cm$^{-1}$). However, we noticed a small change when the M (Na, Li, Ti) atoms are at the cation sites (Cu) in the pure Cu$_2$O. We remark the same peaks when we doped with Na with a small displacement of the second peak (from 800 to 650 nm). The Li or Ti doped Cu$_2$O gives the same result in visible and UV regions but without the initial peak recorded in the pure system, and the absorbance increases between 650 and 750 nm. Furthermore, incorporation of Hydrogen reduces the absorbance of pure Cu$_2$O and Ti doped Cu$_2$O. Hydrogen does not decrease the absorbance of Na or Li-doped Cu$_2$O.
%%%%%%%%%%%%%%%%%%%%%%%%%%%%%
%%%%%%%%%%%%%%%%%%%%%%%%%%%%%%

\section{Conclusion}
We studied the electronic properties of M-doped Cu$_2$O (M= Na, Li, Ti) with and without hydrogen by first-principles calculations. The results show that hydrogen is most stable in the tetrahedral site with the low formation energy. Interstitial hydrogen reduces the band gap value of Cu$_2$O and it gives an n-type conduction behavior. Adding hydrogen does not decrease the band gap value of Na or Li-doped Cu$_2$O but it changes the charges conductivity type from p- to n-type. In these cases, the distance between H atom and Na or Li atom does not affect the structural and electronic properties of Na or Li-doped Cu$_2$O systems. The Ti doping of Cu$_2$O changes its conductivity type. Hydrogen does not change the n-type characteristic of Ti doped Cu$_2$O system (hydrogen effect is negligible compared to that of Ti) but it decreases the magnetic moments of this system.  The distance between H and Ti atoms can influence the structural, electronic and magnetic properties of Ti doped Cu$_2$O. From these results, we see that Na is the best p-type dopant in the Cu$_2$O and it is necessary to remove hydrogen in this material to achieve an increased performance. The absorption shows that Cu$_2$O is a promising optical material in the visible region. In this paper, we established the effects of H, Ti, Li or Na doping, which lead to a significant decrease of the absorption, i.e., increase the optical transmittance. The results of this work make it possible to adapt Cu$_2$O to several technological applications and to different scientific fields. In the next stage of our research, we will study the dopants concentration effects and we hope to confirm our results by experimental investigations.
\section*{Acknowledgements}
This work was funded under the Algerian General Directorate of Scientific Research and Technolo\-gical Development (DGRSDT).

\ukrainianpart
\title{Першопринципне дослідження поведінки водню у M-легованому Cu$_2$O (M $=$ Na, Li і Ti)%
}
\author{A. Лярабі\refaddr{1}, A. Махмуді\refaddr{2}, M. Мебаркі\refaddr{1}, M. Дергал\refaddr{2}}
\addresses{\addr{1}Науково-дослідний центр з технології напівпровідників для енергетики (CRTSE),
	BP 140 Алжир 7- Мервей 16038, Алжир 
	\addr{2} 
	Відділення вивчення та прогнозування матеріалів, відділ досліджень матеріалів та відновлюваних джерел енергії, LEPM-URMER, Університет Тлемсена, Алжир
}
	
	\makeukrtitle
	\begin{abstract}
		Досліджується вплив водню на електронні, магнітні та оптичні властивості Cu$_2$O у присутності різних легуючих домішок (Na, Li і Ti). Розрахунки електронних властивостей показують, що водень змінює провідність Cu$_2$O з p до n-типу. Результати показують, що міжвузловий атом водню воліє розташуватися в тетраедричній ділянці в системі Cu$_2$O і зменшує   ширину забороненої зони останньої. Cu$_2$O,  легований Na або Li зберігає p-тип провідності Cu$_2$O, в той час як водень є джерелом n-типу провідності в системах Cu$_2$O, легованих   Na або Li.   Ti легування збільшує ширину забороненої зони Cu$_2$O і робить її напівпровідником n-типу. Водень підвищує коефіцієнт оптичного пропускання  М-легованого Cu$_2$O.

		\keywords Cu$_2$O, водень, теорія функціоналу густини
	\end{abstract}

\end{document}